\documentclass[pdflatex,sn-mathphys-num]{sn-jnl}

\usepackage{comment} 
\usepackage{tabularx}
\usepackage[utf8]{inputenc}
\usepackage{hyperref}
\usepackage{graphicx}%
\usepackage{multirow}%
\usepackage{amsmath,amssymb,amsfonts}%
\usepackage{amsthm}%
\usepackage{mathrsfs}%
\usepackage[title]{appendix}%
\usepackage{xcolor}%
\usepackage{textcomp}%
\usepackage{textgreek}
\usepackage{manyfoot}%
\usepackage{booktabs}%
\usepackage[utf8]{inputenc}
\usepackage[utf8]{inputenc}
\usepackage{newunicodechar}
\newunicodechar{⁻}{\textsuperscript{-}}
\usepackage{algorithm}%
\usepackage{algorithmicx}%
\usepackage{algpseudocode}%
\usepackage{listings}%

\theoremstyle{thmstyleone}%
\theoremstyle{thmstyletwo}%

\theoremstyle{thmstylethree}%

\usepackage{textcomp}
\usepackage{multibib}
\newcites{SI}{Supplementary References} 
\begin{document}

\title{Exotic topological defects and director fields in free-floating spherical ferroelectric nematic liquid crystal shells}


\author[1]{\fnm{Churchill B.} \sur{Agoni}}

\author[2]{\fnm{Ema} \sur{Pilih}}

\author[2,3]{\fnm{Luka} \sur{Cmok}}

\author[4]{\fnm{Calum J.} \sur{Gibb}}

\author[5]{\fnm{Jordan} \sur{Hobbs}}

\author[4,5]{\fnm{Richard} \sur{Mandle}}

\author[2,3]{\fnm{Irena} \sur{Drevensek-Olenik}}

\author*[1]{\fnm{Jan P.F.} \sur{Lagerwall}}\email{jan.lagerwall@lcsoftmatter.com}

\affil*[1]{\orgdiv{Department of Physics \& Materials Science}, \orgname{University of Luxembourg}, \orgaddress{\street{162a, avenue de la faiencerie}, \city{Luxembourg City}, \postcode{1511}, \country{Luxembourg}}}

\affil[2]{\orgdiv{Faculty of Mathematics and Physics}, \orgname{University of Ljubljana}, \orgaddress{\street{Jadranska 19}, \city{Ljubljana}, \postcode{1000}, \country{Slovenia}}}

\affil[3]{\orgdiv{Department of Complex Matter}, \orgname{J.~Stefan Institute}, \orgaddress{\street{Jamova 39}, \city{Ljubljana}, \postcode{1000}, \country{Slovenia}}}

\affil[4]{\orgdiv{School of Chemistry}, \orgname{University of Leeds}, \city{Leeds}, \country{UK}}

\affil[5]{\orgdiv{School of Physics and Astronomy}, \orgname{University of Leeds}, \city{Leeds}, \country{UK}}


\abstract{Ferroelectric nematic ($N_\mathrm{F}$) liquid crystals exhibit polar symmetry and large polarization, giving rise to phenomena absent in conventional apolar nematics. We investigate $N_\mathrm{F}$ liquid crystals confined to free-floating spherical shells with tangential boundary conditions, enforcing a total topological defect charge of +2. We conjecture that ferroelectric nematics avoid splayed configurations with half-integer defects---common in apolar nematic shells---instead concentrating the topological charge into escaped azimuthal +1 defects requiring only bend and twist. Indeed, at room temperature in the N$_F$ phase, our thin RM734+DIO shells with inner and outer aqueous poly(vinyl alcohol) solutions develop an azimuthal director field around two antipodal +1 bend-twist defects. The non-centrosymmetric nature and the azimuthal director configuration of the shells in the $N_\mathrm{F}$ phase are confirmed also through second-harmonic generation microscopy. At intermediate temperature the antiferroelectric $N_\mathrm{x}$ phase generates a new exotic texture rife in zigzag lines in the shells. In the regular N phase at high temperature, the shells develop the usual four +1/2 disclinations located near the thinnest point. Our study highlights the rich platform offered by spherical shells to study the behavior of exotic liquid crystals subject to topological constraints, possibly opening new paths to apply the highly responsive ferroelectric nematic phase.
}




\maketitle

\section*{Introduction}\label{sec1}

Liquid crystal (LC) shells---each a thin spherical layer of LC confined between an internal aqueous droplet and an external aqueous continuous phase---are convenient model systems for exploring the interplay between topological constraints and long-range orientational order. In LCs such order is described by the director field \textbf{n}(\textbf{r}), which specifies the preferred orientation \textbf{n} of the LC molecules (mesogens) at each spatial position \textbf{r}. When LC order is confined within a closed curved surface with a component of \textbf{n}(\textbf{r}) in the plane of the surface, the formation of topological defects becomes unavoidable: the Poincar\'e--Hopf theorem requires that the total topological charge of \textbf{n}(\textbf{r}) in any such surface equals $s=+2$ \cite{lopez2011drops}. In a conventional nematic LC phase ($N$)---where $\mathbf{n}\equiv -\mathbf{n}$, i.e., the director exhibits head-to-tail invariance---confined to a shell with tangential boundary conditions on the inner and outer interfaces this constraint is typically satisfied by the formation of four $s=+1/2$ disclinations, arranged so that they minimize the Oseen--Frank elastic energy \cite{lopez2011drops,lopez2011frustrated,Urbanski2017a}. This means that defects are formed as to minimize director field bend while favoring splay, since the latter gives a lower elastic energy contribution in conventional $N$ phases, and in asymmetric shells they are collected near the thinnest point to minimize the defect extension within the LC. Studies of LC shells with varying stabilizers, yielding complete or partial tangential alignment \cite{lopez2009topological,Noh2016LCshells}, sometimes changing with temperature \cite{Noh2020a,Sharma2019LCshellRealignment,Ma2024}, have yielded important insights into defect interactions, elastic anisotropy, and curvature-induced frustration in (chiral) nematic liquid crystal systems.

The recent discovery of the ferroelectric nematic ($N_\mathrm{F}$) phase \cite{Chen2020b} has introduced a fundamentally new ingredient into this framework: a spontaneous electric polarization \textbf{P}, arising from the broken head-to-tail invariance of the phase. Due to the vectorial nature of \textbf{P}, defects with half integer topological charge are generally not permitted; if they appear, a costly $\pi$-wall in the \textbf{P}-field must form between them. Additionally, because \textbf{P} is rigidly coupled to the director \textbf{n}, electrostatic interactions are directly transmitted into the elastic free energy \cite{mandle2020ferroelectricity,lavrentovich2021polar}. Consequently, splay distortions, characterized by $\nabla\cdot \mathbf{n}\neq 0$, acquire an additional energetic cost since they generate bound electric charge, $\rho=-\nabla\cdot\mathbf{P}$ \cite{sebastian2023polarization,basnet2025periodic}. The associated electrostatic energy strongly suppresses splay formation, leading to a pronounced splay stiffening. As a result, bend and twist distortions become energetically favored and dominate the textures observed in planar geometries \cite{lavrentovich2025twist,Bennett2025,zavvou2025signatures}. 

While $N_\mathrm{F}$ textures in flat planar geometries are now relatively well understood, their behavior under curved confinement remains largely unexplored. Three studies investigated $N_\mathrm{F}$ growing from the isotropic liquid on cooling, as sessile droplets \cite{Perera2023} or as cylinders connecting the substrates of a standard glass cell \cite{Yang2022, Zou2026}, but freely suspended fully spherical samples have not yet been studied. Confining the $N_\mathrm{F}$ phase to a spherical shell provides a stringent test of how polar order reshapes topological ground states. This raises a central question: how do polarity and the associated electrostatic suppression of splay influence the manner in which the required total topological charge of $s=+2$ is realized?

To investigate this question, we fabricate spherical shells of a ferroelectric nematic LC, prepared as a binary mixture of two commonly studied $N_\mathrm{F}$-forming mesogens, RM734 \cite{Mandle2017} and DIO \cite{Nishikawa2017a} (see Fig.~\ref{fig.structures} for structures), within their ideal-mixing regime \cite{Chen}. Using a solvent-exchange-driven technique we generate stable water-in-oil-in-water core--shell emulsions. The extension of LC shell research to the $N_\mathrm{F}$ phase represents a key novelty of the present work, our polarized optical microscopy (POM) studies revealing a topology markedly different from that of conventional nematic shells. Instead of the usual four +1/2 disclinations with preference for splay, shells in an $N_\mathrm{F}$ phase adopt an azimuthal (in-plane concentric) director field containing two +1 defects surrounded by a splay-free \textbf{n}(\textbf{r}). 
Thermal cycling reveals a reversible topology transformation across the $N \leftrightarrow N_\mathrm{x} \leftrightarrow N_\mathrm{F}$ sequence, where the formation of the intermediate antiferroelectric $N_\mathrm{x}$ phase\footnote{Alternative notations for this phase, which is not in focus for this paper, are $N_s$, $N_{AF}$ or SmZ$_A$} \cite{ghimire2025dynamics} is recognized through a characteristic zigzag texture that is distinct from both $N$ and $N_\mathrm{F}$ textures.


We also perform Second-Harmonic Generation (SHG) microscopy imaging of the shells in the low-temperature phase. These measurements provide direct evidence of broken head-to-tail invariance, supporting the assignment of the phase in the shell as ferroelectric nematic \cite{Nishikawa2017,Sebastian2022}. Polarimetric analysis of the SHG signal further supports an azimuthal director configuration, sometimes with indications of an additional weak spiral modulation suggesting an escaped vortex configuration of polarization field \cite{Yang2022,Zou2024}.

\section*{Results}\label{sec2}
\subsection*{\subsection*{Shell preparation at room temperature and POM texture analysis of $N_\mathrm{F}$ shells}}

In our basic procedure (Fig.~\ref{fig.preparationprocedure}), inspired by the method for making isotropic double emulsions of Wang et al. \cite{Wang2022e}, we produce our shells at room temperature by dispensing droplets of an isotropic quaternary mixture (composition in Table~\ref{Tab:compositions}) of RM734 and DIO dissolved in dichloromethane (DCM) and isopropanol (IPA) as co-solvent into an aqueous solution of poly(vinyl alcohol) (PVA). Basically, the aqueous phase acts as a poor solvent that destabilizes the solution and leads to the nucleation of many droplets of an organic phase (the Ouzo/Pastis effect \cite{Vitale2003, Grillo2003}). The PVA is added since it is well known as an efficient stabilizer for LC shells, normally promoting tangential boundary conditions \cite{Sharma2023}. The exact procedure of shell formation is currently under elucidation and we will report on it separately. Our hypothesis is that the shells form as a result of radial phase separation within the nucleated droplets as the IPA mixes with the aqueous phase, in parallel with the volatile DCM diffusing into the aqueous phase and then evaporating from its surface. This would be a more complex analogue to the droplet phase separation procedure developed by Haase and Brujic which did not contain the volatile component \cite{haase2014tailoring}. The procedure reproducibly yields core–shell droplets consisting of an LC shell (consisting primarily of RM734 + DIO although some IPA and possibly trace amounts of DCM and water may also be present) suspended in and surrounding isotropic phases dominated by aqueous PVA solution. The overall formulations and emulsification/solvent-exchange protocols are described in the Methods section and in the Supplementary Information (SI), Section \ref{S1.1}. 

The highly dynamic process as the organic solvents escape the shell is reflected in a texture that initially changes rapidly, see Supplementary Movie 1. The transition from isotropic to an LC state is easily recognized in POM by the shells becoming bright between crossed polarizers (or changing from red to white-yellow if a $\lambda$-plate is inserted), see Fig.~\ref{fig.firsttransition}, but the early LC phases are difficult to identify with certainty due to the dynamics. About 15--20~s after the first appearance of birefringence a texture begins to stabilize that is characteristic of the $N_\mathrm{F}$ phase; this is the room temperature phase of our mixture. Comparing with later experiments at varying temperature (see below) we conjecture that the full $N\rightarrow N_\mathrm{x} \rightarrow N_\mathrm{F}$ phase sequence is rapidly passed through as the organic solvent content is reduced. 

\begin{figure}[t!]
\centering
\includegraphics[width=0.55\textwidth]{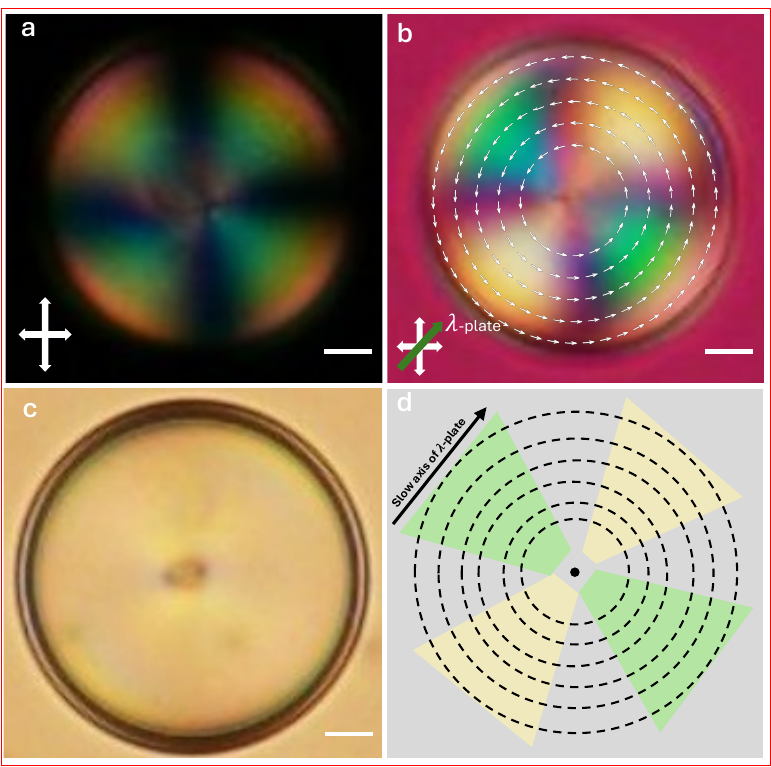}
\caption{\textbf{POM analysis of the director field in $N_\mathrm{F}$ shells}. (a) Between crossed polarizers an $N_\mathrm{F}$ shell exhibits a four-lobed birefringence pattern with a Maltese cross of extinction bands. (b) Inserting a $\lambda$-plate, adjacent lobes alternate in colour, shifting in opposite sense within the Michel-L\'evy chart (Fig.~\ref{fig.detailedPOManalysis} of SI) relative to panel (a), corresponding to locally reduced (white-yellow) and increased (blue-green) effective retardation. (c) The defects are recognized from their scattering character by removing the polarizers. Because the focus is on the equator, both defects are seen simultaneously near the center of the image, slightly out of focus.  (d) The POM textures are consistent with an azimuthal tangential director field that bends everywhere around $s=+1$ defects at the top and bottom, respectively. Scale bar: 10$\mu$m.}\label{fig23}
\end{figure}

Fig.~\ref{fig23}a shows representative POM textures of a stationary equilibrated shell in the $N_\mathrm{F}$ phase, featuring a characteristic four-lobed birefringence pattern with bright sectors separated by mutually orthogonal extinction bands (Maltese cross). This texture is consistent with the presence of antipodal $s=+1$ point defects near the shell's top and bottom, respectively \cite{Noh2016LCshells,Noh2020a}. Due to the action of gravity and the density mismatch between the liquid of the inner droplet (less dense) and the LC (more dense), these correspond to the thinnest and thickest point, respectively, of the shell (see SI Fig.~\ref{fig5a} for a side view micrograph of several $N_\mathrm{F}$ shells at room temperature). The micrograph without polarizers in Fig.~\ref{fig23}c allows us to estimate the thickness at the equator (where this photo is focused) as approximately 1.5~$\mu$m. The equator thickness is slightly larger than the average shell thickness \cite{Sharma2022}, hence we may assume an average shell thickness of about 1.3~$\mu$m.

A detailed analysis of the POM image without and with $\lambda$-plate inserted is provided in Section~\ref{detailedPOM} of the SI. In brief, the analysis confirms that the shells adopt tangential alignment through three key observations. Beyond the two antipodal +1 defects rather than the conventional $N$ shell configuration with all defects close to the thinnest point, we see a circularly symmetric radial color profile that further sets the $N_\mathrm{F}$ shell texture apart. Finally, the color shifts upon introduction of a first-order $\lambda$-plate (Fig.~\ref{fig23}b) clearly confirms an azimuthal director field configuration where \textbf{n}(\textbf{r}) forms concentric circles around each defect. This allows us to sketch the approximate director field of the top (or bottom) shell half mapped onto a plane as in Fig.~\ref{fig23}d.

Right around the center of the shell in Fig.~\ref{fig23}a we notice that the cross is slightly inclined compared to the overall Maltese cross and we see a slightly spiraling connection between the inner and outer regions. As we will argue in the Discussion, we attribute this behavior to the defects having an escaped twisted character \cite{Cladis1972a}, which is translated into a regular azimuthal director field further from the defect. We also note in Fig.~\ref{fig23}c, where both defects are visible (albeit out of focus), that they are not quite on top of each other, and neither is in the exact center of the shell image. This explains the slight deviations from perfect circular symmetry, e.g., slightly different colors in quadrants that would show identical colors had the shell been observed exactly along its symmetry axis.

\begin{figure}[t!]
\centering
\includegraphics[width=0.85\textwidth]{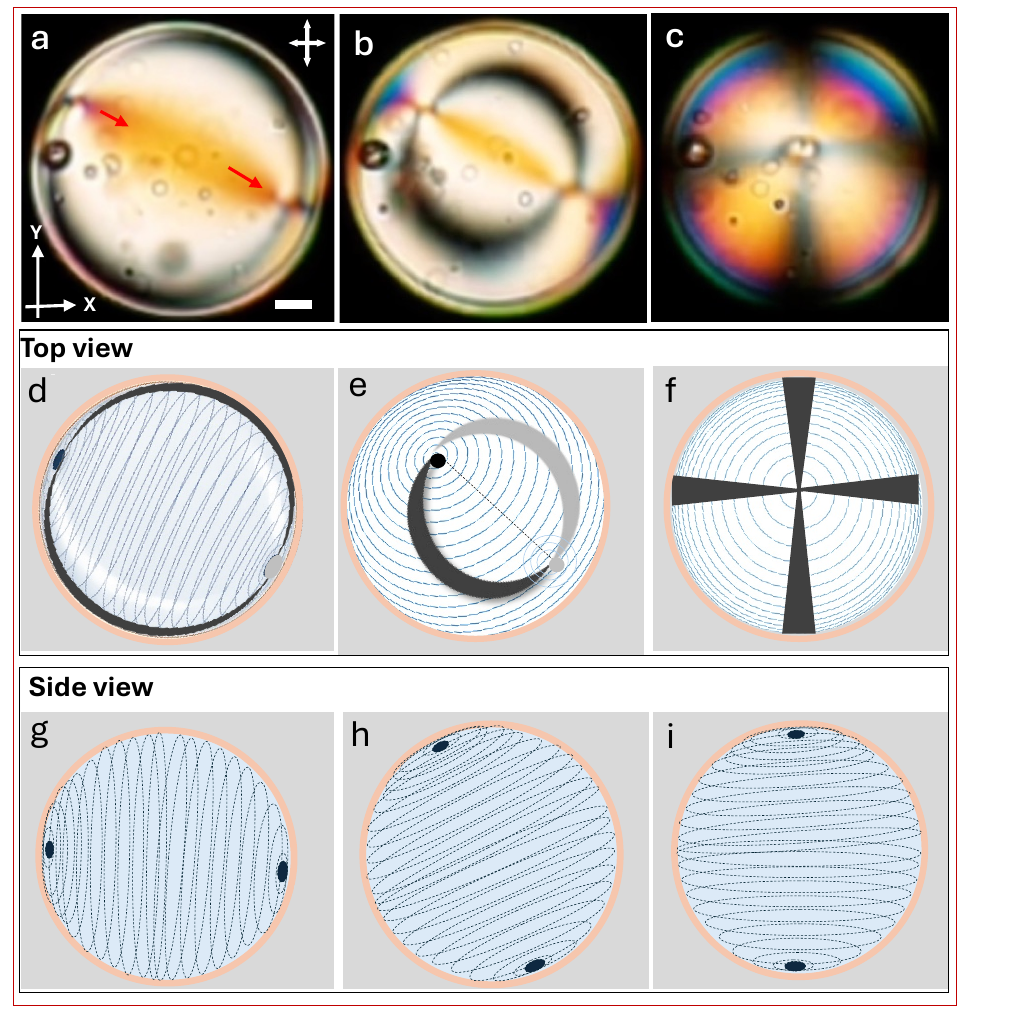}
\caption{\textbf{Sequential POM images showing evolution of shell rotation-induced dynamic texture in an $N_\mathrm{F}$ shell}. (a--b) A dark ring occurs where the projection of \textbf{n}(\textbf{r}) into the imaging plane (d--e) is parallel to one of the polarizers; this ring shrinks during shell rotation, while also the apparent locations of the two defects move inwards parallel/antiparallel to the red arrows in (a), which show the direction of shell translation. (c) Twice per rotation we observe the shell along the axis connecting the two $s=+1$ defects (f), yielding the four-lobed birefringence pattern with a Maltese cross seen also in Fig.~\ref{fig23}a. (g–i) Schematic drawings of \textbf{n}(\textbf{r}) in a side view perspective, i.e., along the shell rotation axis, corresponding to the top view schematics shown in (d--f). Scale bar: 10~$\mu$m.}\label{fig25}
\end{figure}

Sometimes we see shells in the $N_\mathrm{F}$ phase moving with a combination of translation and rotation, either under the influence of gravity or by induced local flows. This motion generates a distinctive periodic modulation of birefringence patterns as observed in POM that are never seen with conventional nematic shells subject to similar motion, see Fig.~\ref{fig25} and Supplementary Movies S2--S4. As the shell undergoes rotation about an axis that is (primarily) in the imaging plane (denoted $x$--$y$) and approximately perpendicular to the axis separating the two defects, a dark ring progressively shrinks inward toward the centre (Fig.~\ref{fig25}a–b) until it almost disappears concurrently with the appearance of a Maltese cross (Fig.~\ref{fig25}c). This sequence, which is coupled to strong local variations in birefringence colour, is then repeated in reverse, and the full cycle recurs with each shell rotation.

As illustrated in the schematic drawings below the micrographs, this texture sequence is another result of the concentric azimuthal director field prevailing in the $N_\mathrm{F}$ shells. The variations in colour and brightness arise from the changing projection of the director field relative to the crossed polarizer axes. As the drawings highlight, the lines of extinction, connecting all locations where \textbf{n}(\textbf{r}) is either parallel or perpendicular to the polarizer or analyzer such that no birefringence effect is seen, vary between one extreme state that is a circle along the perimeter (sketched in (d) and yielding the texture in (a)) and another extreme which is the Maltese cross (f and c). At intermediate angles we see two dark arcs connected into a circle that is always smaller than the shell perimeter, changing size as the shell rotates (e and b). Supplementary Movie S4 compares this type of rotation for shells in the $N_\mathrm{F}$ and $N$ phase, respectively, and it is clear that the textural sequence is entirely different. The regular $N$ shells do not develop the azimuthal director field with concentric rings and therefore the texture displays two hyperbolic arc brushes \cite{Yang2022c} rather than the dark ring seen in the $N_\mathrm{F}$ shells.

\subsection*{POM characterization of shells at different temperatures.} \label{1.2}
To allow the full LC phase sequence of the RM734+DIO mixture to be explored, requiring heating to temperatures approaching the boiling point of water, we modify our initial procedure for making shells by including glycerol. This raises the effective boiling temperature of the isotropic phases and---given that some glycerol most likely dissolves in the LC phase---reduces the intra-LC phase transition temperatures somewhat, thus minimizing evaporation-driven stresses on the shells. The full shell production procedure and the mass compositions of all solutions used in the process are described in Table~\ref{Tab:S3} and the corresponding text in the SI. In short, the injection of the mesogen solution into the PVA solution still takes place at room temperatures, since otherwise the DCM evaporates too quickly, but as soon as the shells have been formed (about a minute after injection) the sample is poured onto  a watch glass and immediately transferred onto an open hot stage placed on the POM that has been pre-heated to $\sim100^\circ$C. The newly formed shells are thus very rapidly heated to the range of the regular $N$ phase, allowing us to study the POM textures of shells of our mixture at temperatures where it develops a conventional $N$ phase or the antiferroelectric $N_\mathrm{x}$ phase, comparing it with the $N_\mathrm{F}$ textures.  

\begin{figure}[b!]
\centering
\includegraphics[width=0.75\textwidth]{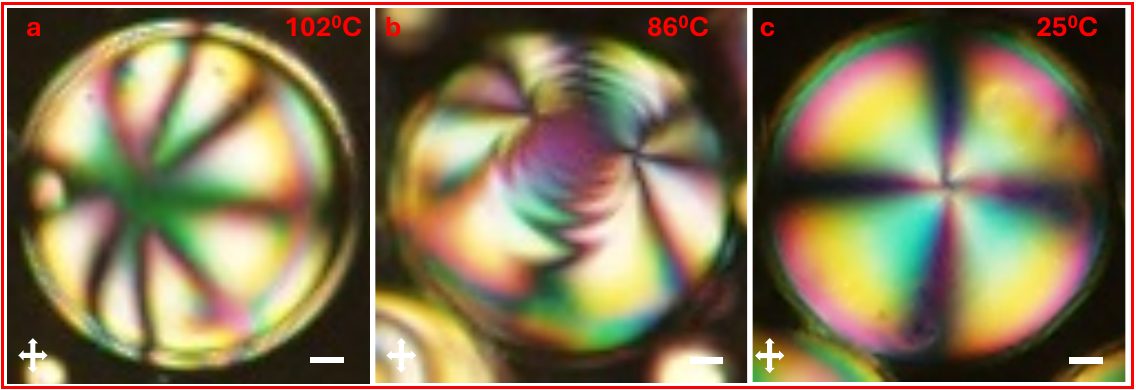}
\caption{\textbf{POM textures of shells in the three different LC phases}. (a) In the high-temperature $N$ phase with +1/2 disclination pairs; (b) after rapid cooling (-10$^\circ$C/min) into the antiferroelectric $N_\mathrm{x}$ phase, transforming the half-integer defects into a network of dark brushes that evolve into pronounced zigzag structures. (c) The shell finally relaxes in the ferroelectric nematic regime and the defect configuration again changes, now into two antipodal +1 defects. Scale bar:10~$\mu$m.}\label{fig24}
\end{figure}

We find two common $N$ shell textures, either with two $s=+1$ defects or with two pairs of $s=+1/2$ disclinations (Fig.~\ref{fig24}a), in both cases near the thinnest point of the shell. As we cool the shells to lower temperatures but without leaving the $N$ phase we note that those with two pairs of $s=+1/2$ disclinations tend to rearrange their director fields such that pairs merge, leaving only shells with two $s=+1$ defects, both close to the thinnest part of the shell. 

In order to study the textures after the transitions to the $N_\mathrm{x}$ and $N_\mathrm{F}$ phases we must cool them rather rapidly past the $N\rightarrow N_\mathrm{x}$ transition, at -10$^\circ$C/min. Slower cooling unfortunately leads to shell rupture at the transition, as previously observed also for tangential-aligned shells of a mixture exhibiting a SmC phase in the sequence when they pass the transition from nematic to smectic order \cite{Sharma2024}. Following this procedure over 60\% of the shells survive the full cooling process, from $N$ through $N_\mathrm{x}$ and finally to $N_\mathrm{F}$. The addition of glycerol into both the modified aqueous phase and the LC/solvent system is essential for enhancing thermal stability. In contrast to conventional PVA/water systems where all shells rupture well below 100$^\circ$C, the glycerol--containing system preserves shell integrity even at temperatures exceeding 100$^\circ$C.

The transition to $N_\mathrm{x}$ is recognized by a highly irregular texture developing (Fig.~\ref{fig24}b) that we have never before seen in our work on LC shells. We see what appears like two partially disrupted $s=+1$ defects separated by a zig-zag-like set of curved extinction lines. It is beyond the scope of this article to analyze this texture, as the $N_\mathrm{x}$ phase is not the focus of our attention, but we note that the texture in Fig.~\ref{fig24}b is representative for shells in this phase (see Supplementary Movie 5 which shows many more examples), allowing its formation to be easily recognized. Subsequent rapid cooling at -10$^\circ$C/min to the $N_\mathrm{F}$ regime results in shells settling with two antipodal +1 defects (Fig.~\ref{fig24}c), as in the first study of $N_\mathrm{F}$ shells that had not been heated. However, comparing the textures in Fig.~\ref{fig24} and Supplementary Movie 1 we note that the dynamic texture sequence during solvent extraction has similarities to that seen on cooling from $N$.

\section*{Optical second harmonic generation (SHG) microscopy in the N$_F$ phase}\label{sec4}
For examining shells by SHG microscopy we follow a room temperature procedure similar to the original one (see Table~\ref{Tab:S3} of SI for details), but the mesogenic materials are synthesized by the co-authors. For this formulation, brief vortex mixing (a few seconds) proved to be most effective for producing relatively large and stable shells. The mixture is then placed on a glass slide with a silicone grease border, and a cover glass is added to seal the sample. When carefully covered to minimize air pockets, the samples remain stable overnight.

An example of a low-magnification SHG microscopy image is provided in the SI Fig.~\ref{figS9}. We first analyze the total SHG signal from individual shells as a function of the incident intensity $I(\omega)$ of the fundamental laser beam. As shown in Fig.~\ref{figS10}, the signal exhibits a quadratic dependence, characteristic of second-order nonlinear optical processes. This behaviour confirms the non-centrosymmetric order of the shells.

\begin{figure}[t!]
\centering
\includegraphics[width=1\textwidth]{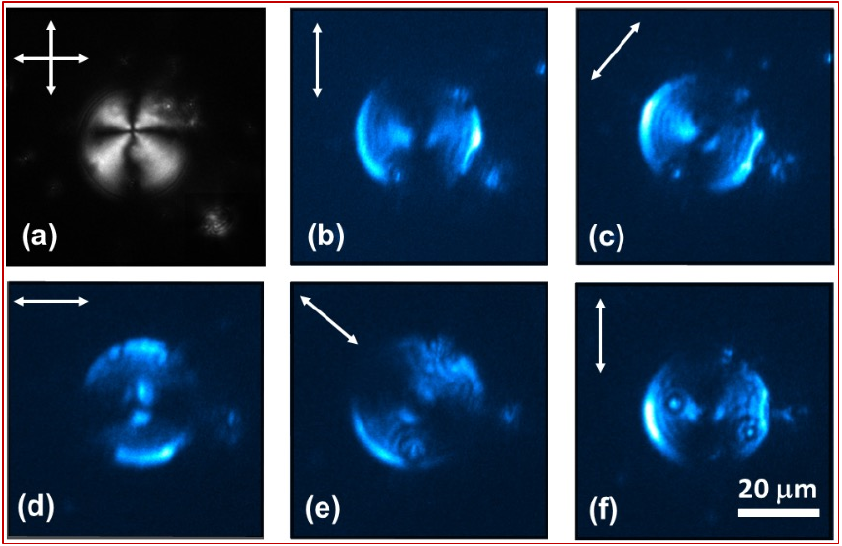}
\caption{\textbf{Second-Harmonic Generation Microscopy characterization of a large $N_{\mathrm{F}}$ shell}. (a) Grey-scale POM image of the shell, with the axis between the two antipodal +1 defects oriented perpendicular to the observation plane, between crossed polarizers obtained using a 405~nm LED as the illumination source. (b)--(f) Wide-field SHG images of the shell, recorded at different orientations of the linearly polarized fundamental beam at 800~nm. The polarization direction is indicated by a double arrow. Images are acquired at 400~nm using a monochrome camera; the coloring is artificial. Scale bar:20~$\mu$m.}\label{fig25a}
\end{figure}

We then select a shell with the axis between the two defects oriented perpendicular to the observation plane and perform SHG microscopy imaging while varying the polarization direction of the linearly polarized fundamental beam. The results are shown in Fig.~\ref{fig25a}b–f. As the incident polarization is rotated, the SHG image, characterized by two bright “wings” located around the central region, rotates accordingly. The orientation of these wings follows the direction of the incident polarization.

The dependence of SHG intensity on the polarization of the fundamental beam is governed by the components of the nonlinear susceptibility tensor $\mathbf{d}$ of the non-centrosymmetric medium. In pure RM734, the component $\mathbf {d}$$_{33}$, with the value around 20~pm/V, strongly predominates \cite{Folcia2022, Lovsin2026}. In contrast, pure DIO has several components of the same magnitude, but their value is much lower (around 0.3~pm/V) \cite{Lovsin2026,Xia2023}. Based on this, it is reasonable to expect that in the 10:90 RM734:DIO mixture used in our experiments, the $\mathbf {d}$$_{33}$ component remains dominant. Consequently, the SHG intensity is maximized when the polarization of the fundamental beam is parallel to \textbf{n}. This confirms that the director field \textbf{n}(\textbf{r}) within the shells is arranged in a concentric manner, in accordance with the schematic drawings shown in Figs.~\ref{fig23} and \ref{fig25}. This result is generally consistent with the observations described in the previous sections. 

\section*{Discussion}\label{sec12}

The defect configurations and director fields observed in these thin $N_\mathbf{F}$ shells offer useful insight into how confinement influences their internal structure. The possibility to form RM734/DIO shells through a process based on the Ouzo/Pastis effect offers a simple, straightforward and reliable way to access these systems, making it easy for anyone to explore the impact of shell confinement of these new LC phases. 

\begin{figure}[b
t!]
\centering
\includegraphics[width=0.5\textwidth]{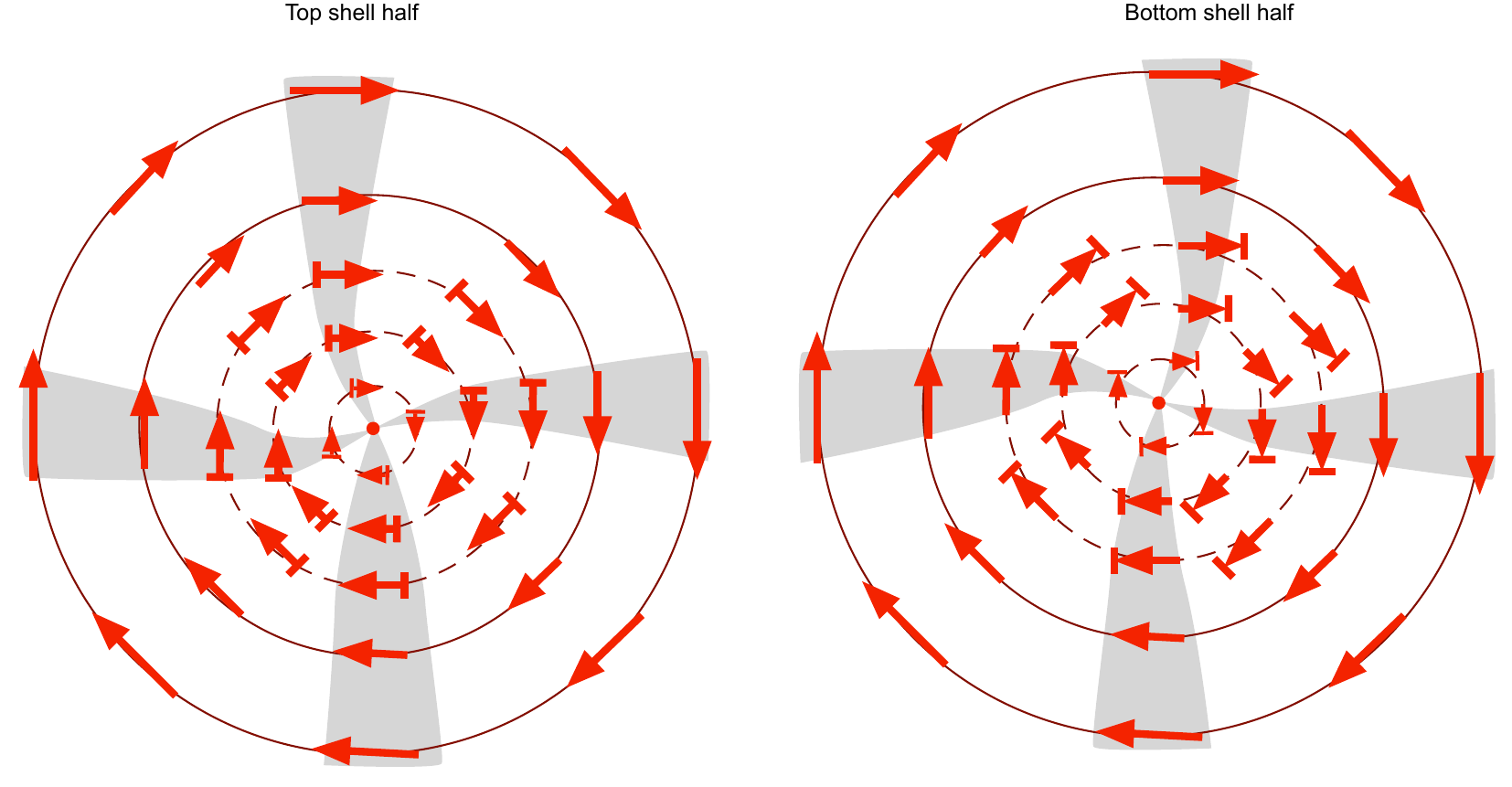}
\caption{\textbf{Escape in the third dimension}. Schematic drawings of how the director field may deform around the top and bottom azimuthal +1 defects of the shell in order to avoid integer defects in the bulk. In both cases the shell is viewed from the top and the red arrows represent the director field inside the LC, between the inner and outer aqueous interfaces. Arrows with a perpendicular end line are tilted out of the screen plane, with the perpendicular line pointing toward us. The highly localized surface charge that must arise at the inner and outer interfaces at each defect, due to the escaped twisted configuration, is most likely compensated by preferential adsorption of negative or positive ions in the water, as appropriate. The arrows are drawn in such a way that they can reproduce the observed subtle spiral extinction crosses observed very near each defect.}\label{EscapeSketches}
\end{figure}

While our analysis gives strong indication that $N_\mathbf{F}$ shells develop antipodal +1 defects at the top and bottom, respectively, we must now consider what happens \textit{within} the shell at each defect location. It is well known that +1 defects are not stable in the bulk, since they 'escape in the third dimension' to reduce the overall free energy \cite{Cladis1972a,Meyer1973}, see Fig.~\ref{EscapeSketches}. In other words, when we speak of a +1 defect at the top or bottom, what we actually have is a +1 surface defect (boojum) at the inside and another at the outside, at the top as well as at the bottom of the shell. There are thus four +1 surface defects, two antipodally on the shell inside and their two respective partners on the shell outside.

The escape in the third dimension happens also in conventional $N$ shells, but their preference for splay means that the director field forms a 'volcano-shaped' splay-bend deformation around each defect. This is not a good option for the $N_\mathbf{F}$ phase, given the high cost of splay, and thus we should rather expect escape via a bend-twist deformation \cite{Cladis1972a}. Such a deformation is often connected to a spiraling pattern in shells to minimize the magnitude of local elastic distortion \cite{Noh2020a}, explaining why we see a tendency to spiraling brushes in the very vicinity of each defect, \textbf{n}(\textbf{r}) transitioning into the concentric circles only after a distance beyond which the escape has no impact, see Fig.~\ref{EscapeSketches}.

Interestingly, during the SHG microscopy experiments we often found shells of much smaller radius which revealed a slight deviation from a concentric director configuration throughout the shell, and the POM textures showed a spiral texture that expanded all the way to the perimeter. The reason why the spiral-like bend-twist escape is sometimes highly localized to the defect and sometimes expands through the entire shell is the topic of on-going investigations and will be reported separately. At present our hypothesis is that the shell-spanning spiral structure is promoted by small radius and/or large average thickness.


In summary, our findings provide clear evidence for an azimuthal arrangement of the director field forming concentric circles around two antipodal +1 defects in $N_\mathrm{F}$ shells. This exotic arrangement is attributed to the intrinsic polarity of the phase, which introduces electrostatic interactions that fundamentally alter the interplay between geometry, topology and elasticity under curved confinement. The polar nature of the N$_F$ phase makes it exceptionally sensitive to external electric fields, opening promising avenues for the controlled manipulation of $N_\mathrm{F}$ shell architectures. We therefore anticipate that the new simple shell preparation method and the described results will stimulate further experimental and theoretical investigations in this emerging field. We expect that electrolytes in the inner and outer aqueous phases may have significant impact on the behavior of \textbf{n}(\textbf{r}) in the shell and thereby on the textures observed. $N_\mathrm{F}$ shells may thus have interesting application areas for instance in sensing applications where conventional $N$ shells do not provide relevant responsiveness. By adding a chiral dopant to turn the $N_\mathrm{F}$ into a ferroelectric cholesteric phase \cite{Feng2021}, the shells can be expected to form highly interesting electric field-responsive optical elements \cite{Lagerwall2026}. Additional open questions include if and how ferroelectric polarization, along with the associated electrostatic interactions, influences the emulsification process and whether this method can be extended to the fabrication of shells from other ferroelectric LC materials.

\section*{Methods}\label{5}
\subsection*{Fabrication of shells}
The LC materials used were two well-known $N_\mathrm{F}$-forming compounds: RM734  and DIO. For the POM studies these were purchased from Instec (USA) and Daken Chemical Ltd. (China), whereas for the SHG measurements the materials were synthesized by the University of Leeds co-authors. The two compounds were mixed at a ratio of 10 wt\% RM734 to 90 wt\% DIO, chosen based on the phase diagram reported in Ref. \cite{Chen}. At room temperature, this mixture exhibits the $N_\mathrm{F}$ phase. The phase transition temperatures were further verified by determining the phase sequence upon cooling, as described in the SI.

For emulsification, the LC compounds were dissolved in dichloromethane (DCM, 99.8\% purity, Sigma-Aldrich) as the primary solvent and isopropanol (IPA, VWR Chemicals) as co-solvent, to introduce the components of what will be the middle phase. The solution was stirred magnetically for approximately 2 minutes at room temperature until a fully isotropic liquid was obtained. The chemical structures of all LC compounds and solvents used in this work and the optimized solvent mass fractions for shell formation are provided in the SI.

A 1 wt\% aqueous solution of poly(vinyl alcohol) (PVA; $M_w = 13–23$~kg/mol; 87–89\% hydrolyzed; Sigma-Aldrich) was used for both the inner and outer isotropic phases. The shells were fabricated via solvent-driven phase separation. In a typical experiment, 100 μL of the LC/DCM/IPA mixture was introduced via pipette into approximately 500 μL of aqueous PVA solution contained in a 2 mL vial. The vial was subsequently gently inverted 2–3 times to ensure uniform dispersion of the LC phase. Immediately after mixing, approximately 100 μL of the resulting suspension was deposited onto a microscopy glass slide and left to equilibrate.

\subsection*{Polarization optical microscopy (POM)}
Polarized optical microscopy (POM) imaging of the shells was performed using an Olympus BX51 microscope equipped with a DP73 camera (Olympus, Japan). With the sample between crossed polarizers, the field of view was initially completely dark, consistent with an optically isotropic state of the solvent-rich shells. To visualize the full shell formation process a full-wave retardation plate ($\lambda$ = 530 nm) was therefore inserted. This enabled the shells to be visualized as black rings against a purple background as they formed during solvent demixing into the continuous phase. After the shells transitioned to the LC state as a result of sufficient solvent removal, the $\lambda$- plate was removed for subsequent standard POM characterization. The kinetics of the shell formation process is shown in Movie 1 of the SI.

\subsection*{Optical second harmonic generation (SHG) microscopy}
SHG microscopy was performed using a custom-made, reconfigurable microscopy setup that enabled the sequential implementation of POM and wide-field SHG microscopy on the same sample. The setup is shown in Fig.~\ref{figS11}. In the SHG configuration, the illumination source was a laser beam from a pulsed Ti: Sapphire laser system (Legend Elite, Coherent), generating light pulses at a wavelength of 800~nm, with a pulse duration of 100~fs and a repetition rate of 1~kHz. The maximum pulse energy was 2.3~$\mu$J. The spot size of a beam on the sample was around 1~cm$^2$.The beam was linearly polarized, and the polarization direction was tuned using a rotating half-wave plate. After appropriate spectral filtering, the 400~nm second-harmonic light generated in the sample was collected with a 20X objective. The corresponding SHG image was recorded using a high sensitivity qCMOS camera (Hamamatsu ORCA-Quest).

For POM characterization during these experiments the setup was reconfigured by exchanging the illumination source to a collimated LED at 405~nm (M405L4, ThorLabs) and inserting the polarizer and analyzer into the optical path. POM images of the shells between crossed and/or parallel polarisers were then captured by using the same objective and camera as for the SHG microscopy. This allowed sequential POM and SHG imaging of the same shell under controlled polarization conditions.

\backmatter

\bmhead{Supplementary information}

Chemical structures and mixture compositions, detailed description of shell preparation procedure, detailed analysis of the POM textures of a typical $N_\mathrm{F}$ shell, details of the SHG setup, and additional texture images and SHG data complementary to those in the main paper.\\ \\
\textit{List of Supplementary Movies}\label{sec13}
\begin{enumerate}
    \item A shell is formed at room temperature as volatile organic solvent evaporates, quickly producing the equilibrium $N_\textrm{F}$ texture with two antipodal $s=+1$ defects surrounded by azimuthal director fields forming concentric rings around each defect.

    \item A shell in the $N_\textrm{F}$ phase translating and rotating such that the texture continuously transitions from a peripheral black ring, gradually shrinking, into a Maltese cross.

    \item The same type of shell rotation as in Movie 2 but starting out without analyzer, allowing the two antipodal defects to be clearly recognized as they rotate with the shell.

    \item Two Movies placed next to one another, showing the rotation of a shell in the $N_\textrm{F}$ phase (left) next to multiple shells in the conventional $N$ phase rotating in a similar way. The texture evolution is completely different in the two cases, reflecting the azimuthal director field in $N_\textrm{F}$ shells which is never seen in $N$ shells.

    \item A large number of shells were initially prepared at room temperature while the organic solvent evaporated slowly. Approximately 1 min after fabrication, before complete solvent evaporation, the shells were transferred onto an open hot stage maintained at $T \approx 120^\circ$C (see SI Movie 6 for isotropic shells transferred onto hot stage). The remaining solvent evaporation was completed on the hot stage, after which the shells exhibited the conventional high-temperature nematic ($N$) phase. Upon cooling, a highly irregular texture emerged in shells that remained intact, indicating the transition into the $N_\mathrm{x}$ phase. At lower temperatures, the texture became smooth again as the shells entered the ferroelectric nematic ($N_\textrm{F}$) phase
    \item Isotropic shells were fabricated at room temperature and immediately transferred onto an open hot stage maintained at $T \approx 120^\circ$C, where they were kept until a single shell in camera view transitioned into the nematic ($N$) phase.

\end{enumerate}

\section*{Declarations}

\subsection*{Funding} 
We acknowledge the financial support of the Slovenian Research and Innovation Agency (ARIS) and the Luxembourg National Research Fund (FNR) within the framework of research project N1-0336 (ARIS) / C23/MS/18125820/SHADOW (FNR) (for the purpose of open access, the author has applied a Creative Commons Attribution 4.0 International (CCBY4.0) license to any Author Accepted Manuscript version arising from this submission). IDO acknowledges support from ARIS within the research programme P1-0192. RM thanks UKRI for funding via a Future Leaders Fellowship, grant number MR/W006391/1.

\subsection*{Competing Interests} 
The authors declare no competing interests.

\subsection*{Data availability} 
All raw data used for this manuscript will be made openly available in a repository on zenodo.org upon acceptance of the manuscript 

\subsection*{Author Contributions}
CBA developed the new method for making shells, carried out all main POM experiments, and came up with the explanation for the textural variations during shell rotation. EP carried out all SHG microscopy and the SHG-related POM experiments and LC assembled the SHG setup. CG, JH, and RM performed chemical synthesis and characterisation of the RM734 and DIO samples used for SHG microscopy. IDO and JPFL developed the original concept and supervised the overall research. CBA, EP, IDO and JPFL wrote the manuscript.

\bibliography{sn-bibliography}

\clearpage
\textbf{\Large{Supplementary Information}}

\setcounter{figure}{0}
\setcounter{table}{0}
\setcounter{section}{0}
\renewcommand{\thefigure}{S\arabic{figure}}
\renewcommand{\thetable}{S\arabic{table}}
\renewcommand{\thesection}{S\arabic{section}}
\section{Materials}\label{S1}
The following materials are used for this study:
\begin{itemize}
    \item \text{4-[(4-nitrophenoxy)carbonyl]phenyl 2,4-dimethoxybenzoate} (RM734).
    \item  \text{4-[(4-nitrophenoxy)carbonyl]phenyl 4-(4-octyloxybenzoyloxy)benzoate} (DIO)
    \item Dichloromethane (DCM)
    \item Isopropanol (IPA)
    \item Polyvinyl alcohol (PVA)
    \item Glycerol
    \item Water 
\end{itemize}
The chemical structures of these compounds and their phase sequences are shown below in Fig.~\ref{fig.structures} and the optimized mass ratio of the mixture used for the production of the ferroelectric nematic shells as well as the phase sequence of this optimized mixture obtained in the cooling direction is given in Table~\ref{Tab:compositions} below.
\begin{figure}[h!]
\centering
\includegraphics[width=0.85\textwidth]{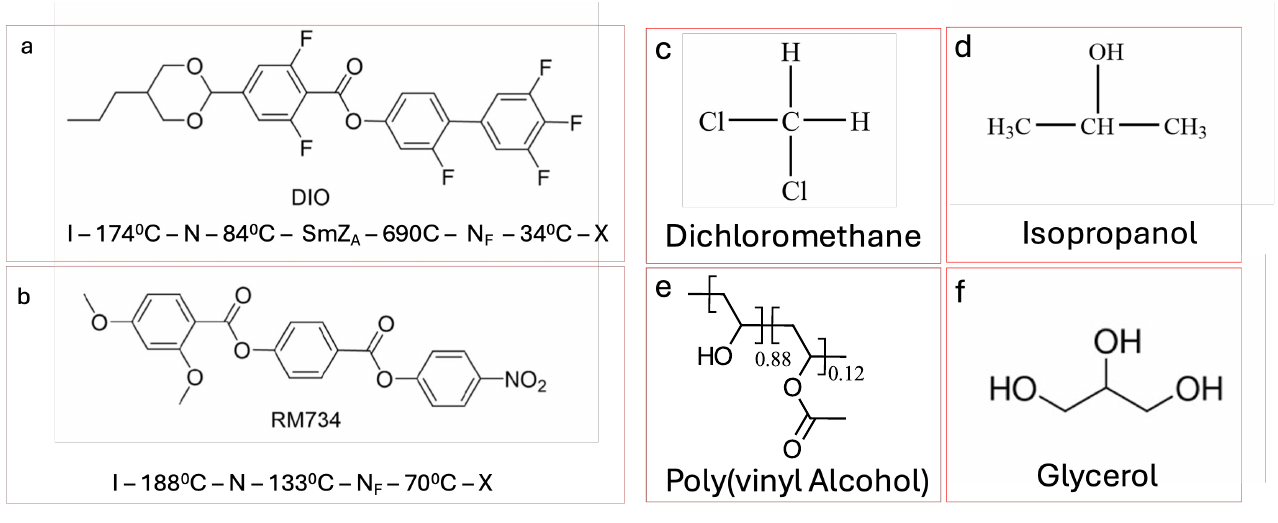}
\caption{Chemical structures and the phase sequences of (a) DIO and (b) RM734 \cite{Chen}. Chemical structure of (c) dichloromethane (DCM) (d) isopropanol (IPA), (e) polyvinyl alcohol (PVA) with 88\% degree of hydrolysis, the remaining $\sim12$\% being polyvinyl acetate from which the PVA is synthesized, and (f) glycerol} \label{fig.structures}
\end{figure}

\begin{table} [h!]
  \centering
  \caption{Optimized mass composition of the two liquid crystal forming molecules, the solvents, the weight and molar concentration of the used PVA and the phase sequence of the mixture on cooling }
  \label{Tab:compositions}
  \begin{tabular}{|l|c|r|}
    \hline
    Compound & Mass (g)&Composition by mass (\%)  \\
    \hline
    RM734 & 0.0023 & 0.44\\
    DIO & 0.0216 & 4.11 \\
    DCM & 0.2660 & 50.58 \\
    IPA & 0.2360 & 44.87 \\
    \hline
    \hline 
    Phase sequence after solvent removal & - & I -- 174$^\circ$C -- $N$ -- 90$^\circ$C -- $N_\mathrm{x}$--75$^\circ$C -- $N_\mathrm{F}$ -- $\sim24^\circ$C -- X\\
    \hline
    \hline 
    \hline 
    PVA & - & 1 wt\%,  13--23kg/mol, 87-89\% hydrolyzed\\
    \hline
  \end{tabular}
\end{table}

\newpage
\subsection{Formation of ferroelectric nematic LC shells}\label{S1.1}
\begin{figure}[h!]
\centering
\includegraphics[width=0.75\textwidth]{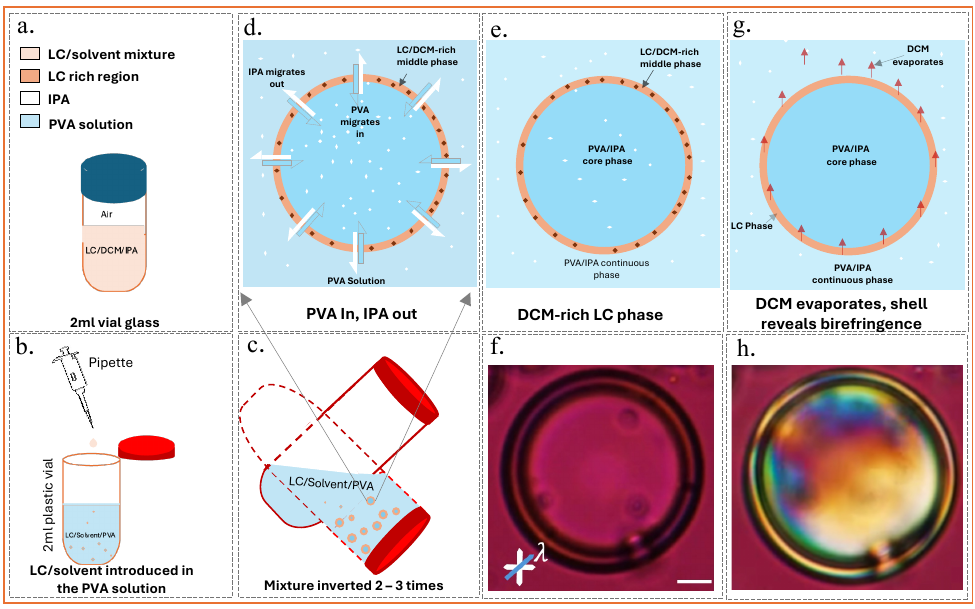}
\caption{Schematic of shell formation via solvent-exchange in an LC/solvent droplet dispensed in aqueous PVA and snapshots of transmission images from polarised optical microscopy (POM) with a 530~nm~$\lambda$‑plate (see  Supplementary Movie 1). (a) An initially homogeneous isotropic LC/solvent mixture (LC + DCM + IPA) with a colour-coding legend left. (b) The mixture is dispensed into PVA solution with a pipette. (c) The mixture is inverted 2 to 3 times to disperse the LC/solvents in PVA solution. (d) Schematic of the mixture emulsified in the PVA–water continuous phase, where IPA migrates into the continuous phase (white arrows), which favours the inflow of PVA solution (blue arrows) into the core phase that pushes the formation and stabilization of double emulsion. (e--f) Schematic and POM image of DCM-rich LC phase view with a full-wave $\lambda$--plate (530~nm) (g--h) Following substantial DCM evaporation and relaxation of solvent gradients, a stable LC shell surrounding an aqueous core remains. Pronounced birefringence confirms the formation of a continuous LC shell with planar anchoring imposed by the PVA-decorated interface. Scale bar: 20~$\mu$m}\label{fig.preparationprocedure}
\end{figure}

\begin{figure}[h!]
\centering
\includegraphics[width=0.7\textwidth]{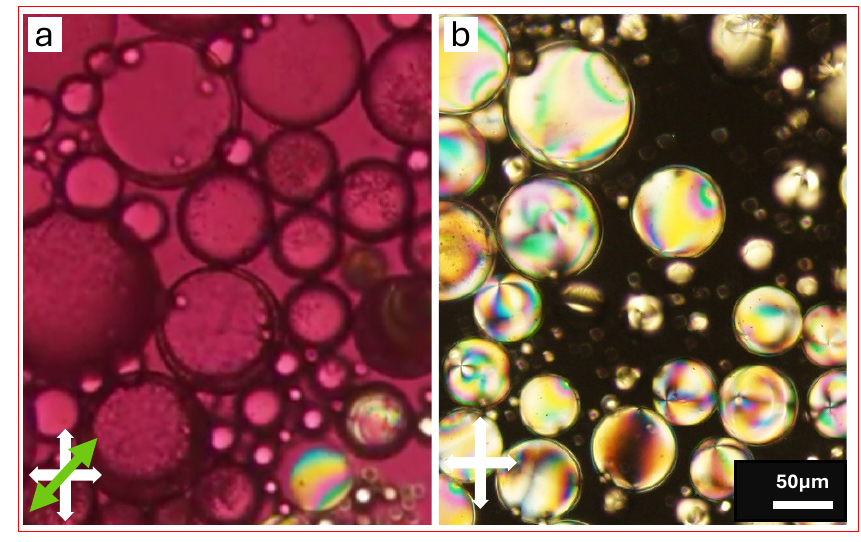}
\caption{POM image of N$_F$ shells formed via a solvent-exchange-induced technique for LC/solvent dispensed in an aqueous PVA continuous phase. (a) A region with isotropic shells rich in DCM solvent. (b) Stable N$_F$ shells with pronounced birefringence formed as DCM completely evaporated from the mixture.}\label{fig.firsttransition}
\end{figure}

 The ferroelectric nematic ($N_\mathrm{F}$) LC shells were fabricated through a solvent-exchange process, transitioning from an initially optically isotropic state to a birefringent shell structure. The homogeneous RM734/DIO solution in DCM/IPA (Fig.~\ref{fig.preparationprocedure}a), when dispensed into the aqueous PVA continuous phase, forms LC/solvent droplets with no birefringent texture (Fig.~\ref{fig.preparationprocedure}b-c). IPA, owing to its high miscibility with water, rapidly diffuses from the droplets into the surrounding aqueous PVA phase. The resulting solvent exchange induces penetration of the PVA solution into the droplet core, leading to the formation of a water-in-oil-in-water (W/O/W) emulsion (Fig.~\ref{fig.preparationprocedure}d). In contrast, DCM, which is  immiscible with water, remains predominantly within the middle phase (Fig.~\ref{fig.preparationprocedure}e–f). In the microscope with the full--wave plate inserted at 45$^\circ$ to the crossed polarizers the shell appears with patches at the droplet periphery. The DCM gradually diffuses through the surrounding continuous phase and evaporates (Fig.~\ref{fig.preparationprocedure}g–h). This solvent exchange progressively enriches the droplets in RM734/DIO and depletes them of IPA and DCM, thereby shifting the local composition across the coexistence boundary of the mixture \citeSI{Reyes2019,NFLC4} into the thermodynamically stable ferroelectric nematic LC shells. Images of multiple shells, first in the isotropic and then in the liquid crystalline state, are shown in Fig.~\ref{fig.firsttransition}. 
 
While the droplets are generally small, with diameters on the order of a few micrometres, the shells exhibit diameters in the range of 20~$\mu$m to 120~$\mu$m. They remain stable for several hours but over longer timescales they rupture and transform into droplets. A systematic variation of the solvent ratio reveals that the shell geometry (diameter and thickness) depends strongly on the IPA concentration in the mixture. A detailed analysis of this effect will be presented in future work.

\clearpage

\section{Gravity-- and density--induced shell thickness variation}

\begin{figure}[h!]
\centering
\includegraphics[width=0.75\textwidth]{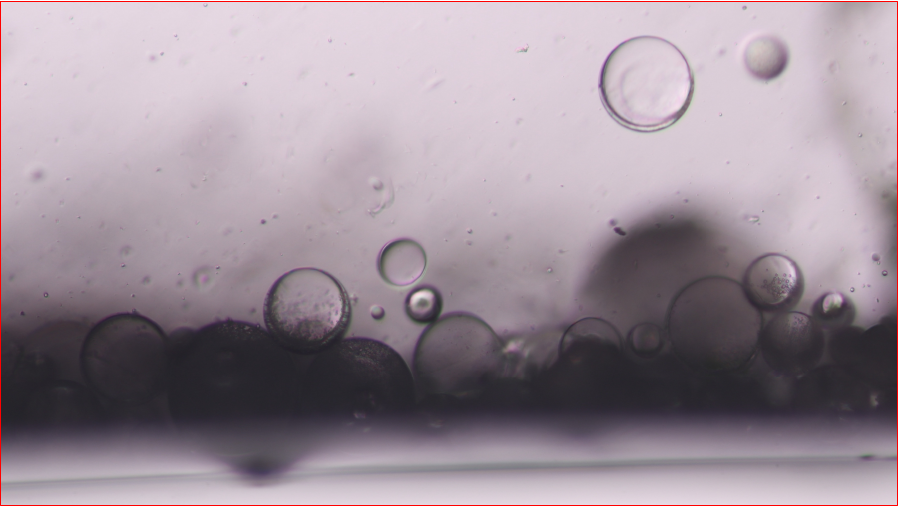}
\caption{Side-view micrograph of several $N_\mathrm{F}$ shells at room temperature, showing the shell thickness variation arising from gravity and the density mismatch between the less-dense inner droplet liquid and the more-dense LC.
}\label{fig5a}
\end{figure}

To image shells from the side, i.e., perpendicular rather than along gravity, we filled the shells into a rectangular cross section glass capillary (400~$\mu$m thickness),  fixed it to an xy-translator stage on the microscope stage and then tilted the entire setup 90$^\circ$. The camera is fixed on top of the microscope so it continuously monitors the sample during the rotation. The rotation of the setup causes the shells to sink down due to the greater density of the LC phase than the continuous phase and we must thus translate the sample until the bottom edge of the capillary, beyond which the shells are unable to move, and there we see how the shells are clearly thinnest at the top and thickest at the bottom, see Fig.~\ref{fig5a}. 

\newpage
\section{Detailed POM analysis of shell birefringence to elucidate the director field configuration}\label{detailedPOM}

\begin{figure}[b!]
\centering
\includegraphics[width=1\textwidth]{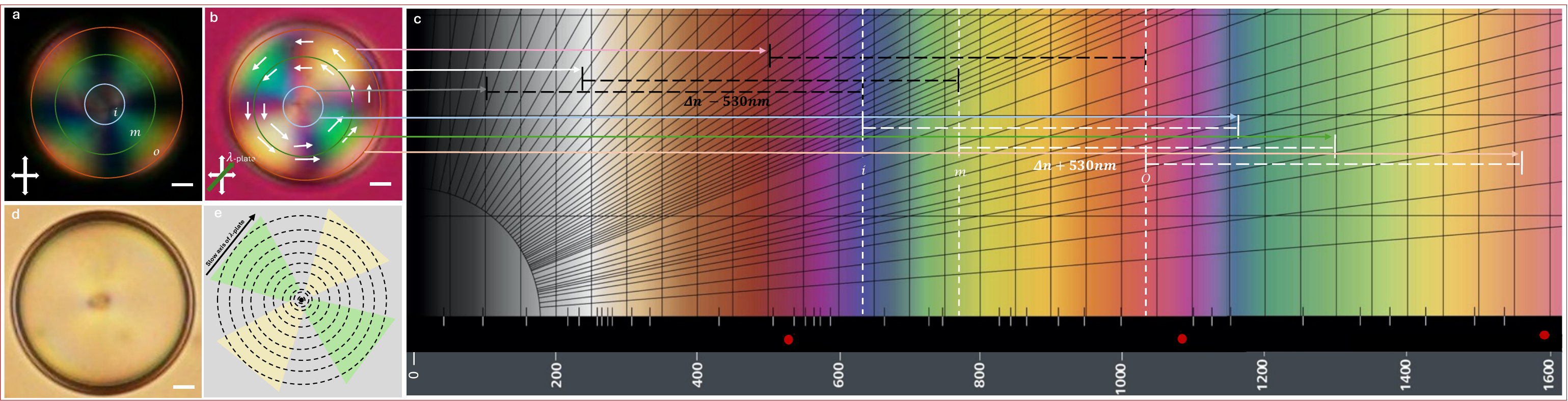}
\caption{\textbf{POM analysis of the director field in $N_\mathrm{F}$ shells}. (a) Between crossed polarizers an $N_\mathrm{F}$ shell exhibits a four-lobed birefringence pattern with a Maltese cross of extinction bands. (b) Inserting a $\lambda$-plate, adjacent lobes alternate in colour, shifting in opposite sense within the Michel-L\'evy chart (c) relative to panel (a), corresponding to locally reduced and enhanced effective retardation. Yellow lobes exhibit reduced effective retardation while green lobes have increased effective retardation. The texture is consistent with an azimuthal tangential director field that bends everywhere around $s=+1$ defects at the top and bottom, respectively. (d) The defects are recognized by their scattering character by removing the polarizers. Because the focus is on the equator, both defects are seen simultaneously near the center of the image, slightly out of focus. (e) Compiling the evidence from the three micrographs we can conclude that the shell exhibits an azimuthal director field where in any spherical plane \textbf{n}(\textbf{r}) circles around the $s=+1$ defects at the top and bottom, respectively, as shown schematically. Scale bar 10$\mu$m. Panel (c) is reproduced on CC-BY license from https://en.wikipedia.org/wiki/Michel-Levy\_interference\_color\_chart.}\label{fig.detailedPOManalysis}
\end{figure}

Fig.~\ref{fig.detailedPOManalysis} shows a more detailed analysis of the POM images of an $N_\mathrm{F}$ shell shown in Fig.~\ref{fig23} in the main paper. We see a pronounced difference in the color profile that sets $N_\mathrm{F}$ shell textures apart from those of conventional $N$ shells. The latter do not have circular symmetry but rather nearly constant color along one direction in the imaging plane, while along the orthogonal direction it changes towards higher retardation in the Michel-Lévy chart from the center to the perimeter. As explained in Appendix A of Noh et al. \cite{Noh2020a}, this reflects the fact that \textbf{n} is oriented normal to the perimeter along the constant-color direction while it is oriented parallel to the perimeter along the orthogonal direction. Upon moving out from the center along the former direction, the effective birefringence continuously decreases by a factor $\cos{\theta}$, where $\theta$ is the inclination angle of the shell wall relative to the optical axis of the microscope, since \textbf{n} is reorienting by this angle towards the orientation of light transmission. At the same time, the increasing inclination of the shell wall yields an increased effective light path through the shell by $2t/cos\theta$, hence the two effects cancel each other out, explaining the constant color. Along the orthogonal direction, \textbf{n} remains in the imaging plane and we maintain the full effective birefringence. The shift in birefringence color along this direction thus fully reflects the increasing effective path difference as the shell wall inclination increases. 

In the $N_\mathbf{F}$ shells, in contrast, the latter behavior prevails around the entire shell, which appears with near-perfect circular symmetry. In other words, there appears to be no reduction in effective birefringence at any point and we only see the impact of increasing path length through the shell as we move from the center towards the perimeter. This is consistent with an azimuthal \textbf{n}(\textbf{r}) forming concentric circles around each defect, as conjectured. Moreover, the color variation from center to perimeter is quite large, ranging from blue close to the center to orange-pink near the perimeter. Taking the birefringence of the DIO/RM734 mixture in the N$_F$ phase to be $\Delta n\approx 0.21$ \citeSI{Song2025a}, the expected optical path difference would be, given that \textbf{n} remains in the image plane thanks to the azimuthal director field, $\Delta L \approx \Delta n\cdot2t/cos\theta$, which yields a minimum value of about 550~nm, corresponding to purple color in the Michel-Lévy chart, for $\theta=0$.

Right around the center of the shell in Fig.~\ref{fig.detailedPOManalysis}a we notice that the cross is slightly inclined compared to the overall Maltese cross and we see a slightly spiraling connection between the inner and outer regions. As we argue in the Discussion, we attribute this behavior to the defects having an escaped twisted character \cite{Cladis1972a}, which is translated into a regular azimuthal director field further from the defect. The twist close to the center means that we cannot apply the Michel-Lévy chart (which assumes uniform optic axis along the viewing direction) in the close vicinity of the defect. Given the $\cos\theta$ factor gaining importance as we move from the center towards the perimeter, we should thus expect the first distinct color, which is blue, to be in the first order, i.e. $L\approx 630$~nm. The orange-pink color towards the perimeter would then correspond to $L\approx 1050$~nm

Insertion of a $\lambda$--plate (530~nm) with its slow axis at 45$^\circ$ with respect to the crossed polarizers produces pronounced alternating birefringence--colour shifts on the shell surface which further confirms the azimuthal director field. The top right and bottom left sectors appear with colors ranging from white close to the defect, over yellow at mid distance, to orange-red near the shell perimeter, while the colors of the top left and bottom right sectors range from blue near the center via green at mid range, shifting to yellow, orange and eventually pink at the perimeter. This is the classic signature of reduced effective retardation in the former case and increased retardation in the latter case.

We draw three circles concentric with the shell perimeter in Fig.~\ref{fig.detailedPOManalysis}a--b to identify inner, mid and outer ranges from the center. We indicate the corresponding retardation colors in the Michel-Lévy chart with the letters $i$, $m$ and $o$, and then draw the addition and subtraction by the $\lambda$-plate of 530~nm from each of these locations to find the expected corresponding colors in panel (b). We draw arrows from that panel to the respective locations in the Michel-Lévy chart, finding good match for added retardation while the micrograph gives a yellow touch in the quadrants of subtraction compared to the Michel-Lévy chart. This is most likely due to imperfect white balance of the microscope--camera system when the micrograph was obtained. 

Given that the $\lambda$-plate slow axis is oriented from bottom left to top right, the results are clear confirmations of an azimuthal distribution of the director field with \textbf{n}(\textbf{r}) forming concentric rings around each defect in the surface of the spherical shell, except in the closest vicinity of the defect where the colors become non-compliant with the Michel-Lévy chart due to the escaped twisted character of \textbf{n}(\textbf{r}) near the defect core. We also note in Fig.~\ref{fig.detailedPOManalysis}d, where both defects are visible (albeit out of focus), that they are not quite on top of each other, and neither is in the exact center of the shell image. This explains the slight deviations from perfect circular symmetry, e.g., slightly different colors in quadrants that would show identical colors had the shell been observed exactly along its symmetry axis.

\section{Topological defects and director fields of \texorpdfstring{$N_\mathrm{F}$}{NF} LC shells}

\begin{figure}[h!]
\centering
\includegraphics[width=0.8\textwidth]{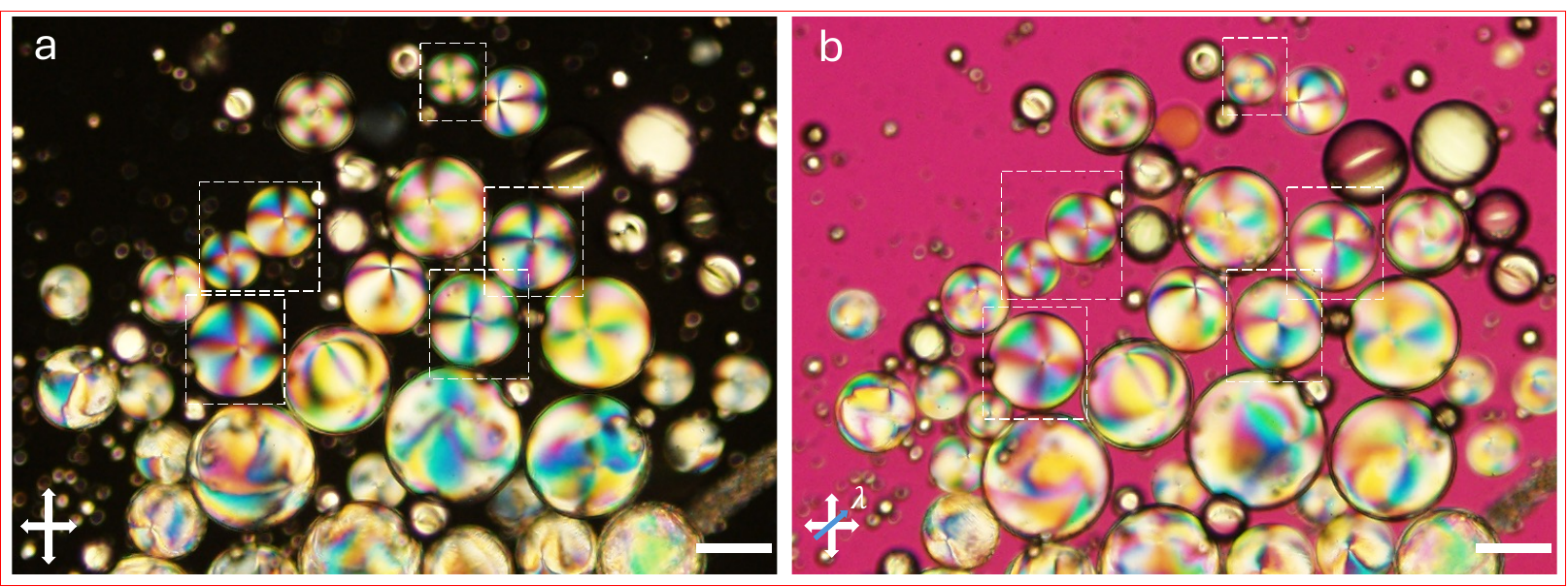}
\caption{POM images of dispersed N$_F$ liquid-crystal shells: (a) Shells between crossed polarizers, many of them in an orientation where they show a four-lobed birefringence pattern with two orthogonal extinction bands with two +1 point defects. (b) The same shells imaged with a $\lambda$ ($530,\mathrm{nm}$) full-wave plate at $45^\circ$ to the polarizers, showing alternating colour shifts corresponding to reduced (yellow) and enhanced (green) effective retardation, consistent with a rotating azimuthal director field.
. Scale bar 50$\mu$m.}\label{fig5}
\end{figure}

POM shell textures from a sample with many shells in the $N_\mathrm{F}$ state, with varying orientations, are shown in Fig.~\ref{fig5}, without (a) and with (b) a first-order $\lambda$-plate inserted in the POM.

\section{Shell preparation for studying textures at varying temperatures}
To observe the textures of shells as they are rapidly heated to the range where the conventional $N$ phase develops and then cooled through the (monotropic) liquid crystal phase sequence, a five-component mixture (RM734+DIO in DCM/IPA/glycerol) was dispensed into a PVA/glycerol/water continuous phase, see Table.~\ref{Tab:S5} and Supplementary Movie 3. The shells were fabricated by dispensing the LC mixture into the glycerol-enhanced PVA solution. The presence of glycerol in both the  aqueous PVA solution and LC/solvent mixture was ideal to elevate the effective boiling point. 

\begin{table} [h!]
  \centering
  \caption{Optimized composition of the five-component mixture, their masses and percentages as well as the mass  of the continuous phase used for the production of the robust N$_F$ shells for the temperature ramp experiment}
  \label{Tab:S5}
  \begin{tabular}{|l|c|r|}
    \hline
    Compound & Mass (g)&Composition by mass (\%)  \\
    \hline
    RM734 & 0.0024 & 0.27\\
    DIO & 0.0217 & 2.43 \\
    Glycerol & 0.2100 & 23.51 \\
    DCM & 0.2660 & 29.78 \\
    IPA & 0.3930 & 44.00 \\
    \hline 
    \hline
    PVA & 1.0000 & 1 wt\%,  13--23kg/mol, 87-89\% hydrolyzed\\
    Glycerol & 20.0000 & 20 wt\% \\
    Water & 79.0000 & 79 wt\% \\
    \hline
  \end{tabular}
\end{table}

\begin{figure}[h!]
\centering
\includegraphics[width=0.75\textwidth]{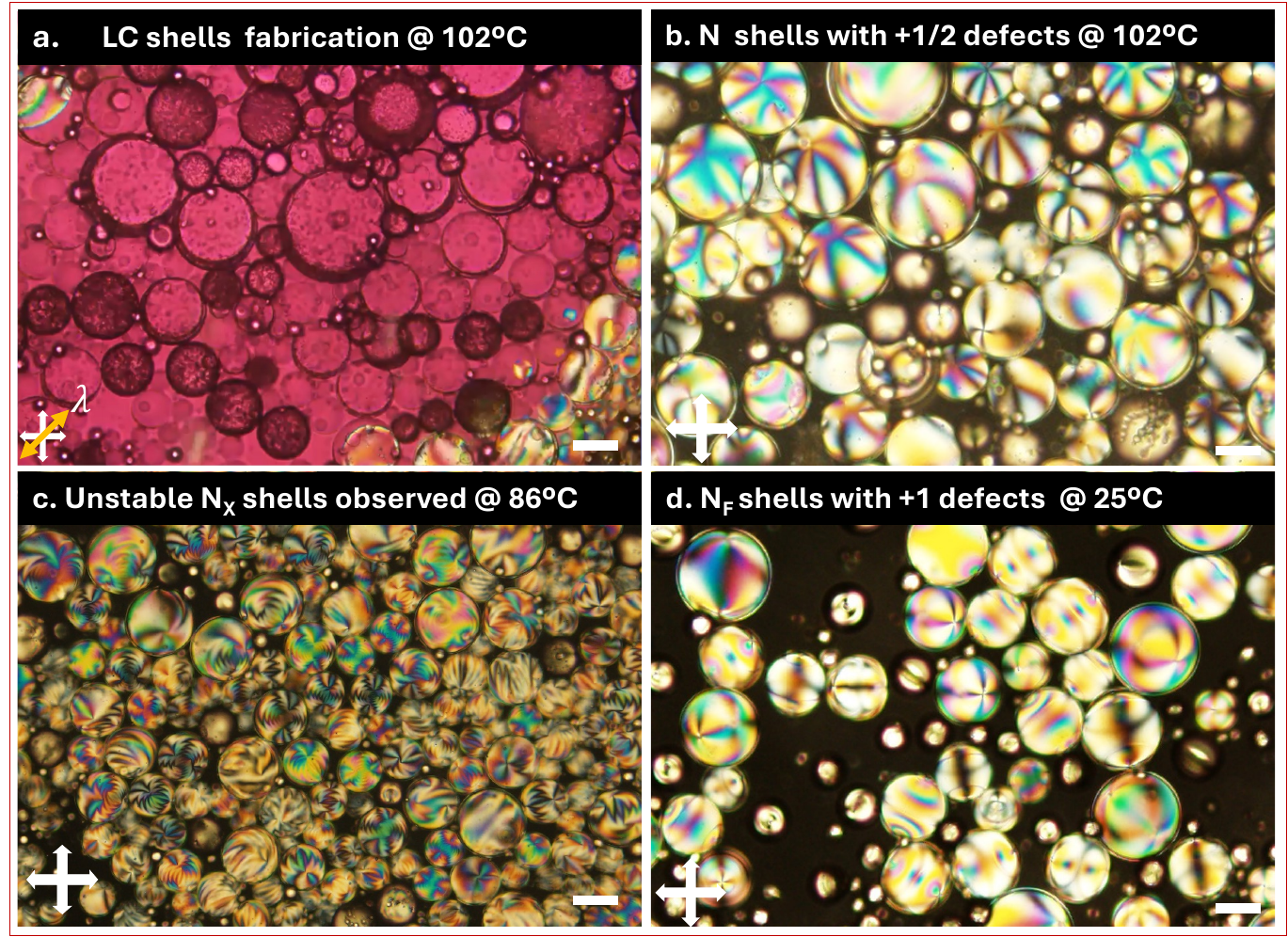}
\caption{POM images of (a) dispersed isotropic solvent-rich LC shells in PVA solution introduced at elevated temperature, (b) nematic shells at high temperature with the characteristic +1/2 defects, (c) shells transiting through the $N_\mathrm{x}$ phase upon cooling, forming a characteristic zigzag texture distinct of this phase, and (d) room temperature ferroelectric nematic LC shells. Scale bar: 50~$\mu$m}\label{figS8}
\end{figure}

\clearpage

\section*{SHG microscopy}
DIO and RM734, used in samples for SHG microscopy, were synthesized in the group of Richard Mandle at the University of Leeds (UK). The compounds were mixed in a weight ratio of RM734 (10~wt\%) to DIO (90~wt\%), forming a liquid crystal (LC) mixture. The LC mixture was then dissolved in a solvent consisting of dichloromethane/isopropanol (DCM/IPA) in a 3:5 ratio, using the proportion 0.005~g LC in 100~μL of solvent (S = DCM/IPA). Subsequently, 40~$\mu$l of this LC–solvent (LCS) mixture was added to 100~$\mu$l of an aqueous 1~wt\% poly(vinyl alcohol) (PVA) solution (see Table~\ref{Tab:S3}).

\begin{table} [h!]
  \centering
  \caption{Mass composition of the three component mixture and the mass composition of the continuous phase for production of $N_\mathrm{F}$ shells for SHG characterization.}
  \label{Tab:S3}
  \begin{tabular}{|l|c|r|}
    \hline
    Compound & Mass (g)&Composition by mass (\%)  \\
    \hline
    LC mixture & 0.0025 & 4.8\\
    DCM & 0.025 & 47.9 \\
    IPA & 0.024 & 47.3 \\
    \hline
    \hline 
    Phase sequence after solvent removal & - & I -- 174$^\circ$C -- $N$ -- 90$^\circ$C -- $N_\mathrm{x}$--75$^\circ$C -- $N_\mathrm{F}$ -- $\sim24^\circ$C -- X\\
    \hline
    \hline 
    \hline 
    PVA & 0.05 & 1 wt\%,  13--23kg/mol, 87-89\% hydrolyzed\\
    water & 49.5 & 99\% \\
    \hline
  \end{tabular}
\end{table}

\begin{figure}[h!]
\centering
\includegraphics[width=1\textwidth]{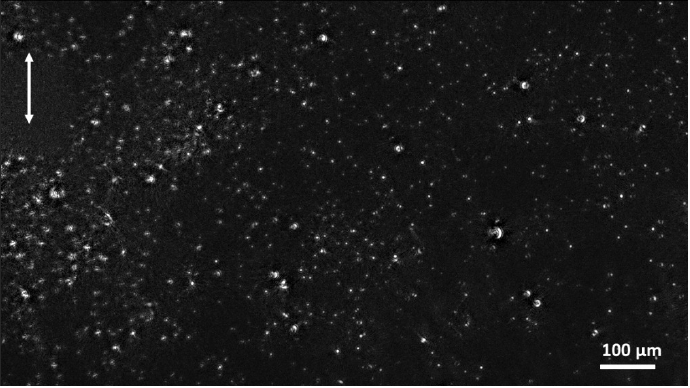}
\caption{Low-magnification, wide-field SHG microscopy image of the suspension. The double arrow indicates the polarization direction of the linearly polarized fundamental beam (800 nm). The image was acquired at the second-harmonic wavelength of 400 nm. Larger structures correspond to shells, while the numerous smaller spots represent droplets. Scale bar: 100~$\mu$m.}\label{figS9}
\end{figure}

The resulting mixture was mechanically agitated for several seconds to induce emulsification. The emulsion was then deposited onto a microscopy glass slide prepared with a silicone grease border. A cover glass was placed on top to seal the sample. Fig.~\ref{figS9} shows the SHG microscopy image of the prepared sample.

\begin{figure}[h!]
\centering
\includegraphics[width=0.95\textwidth]{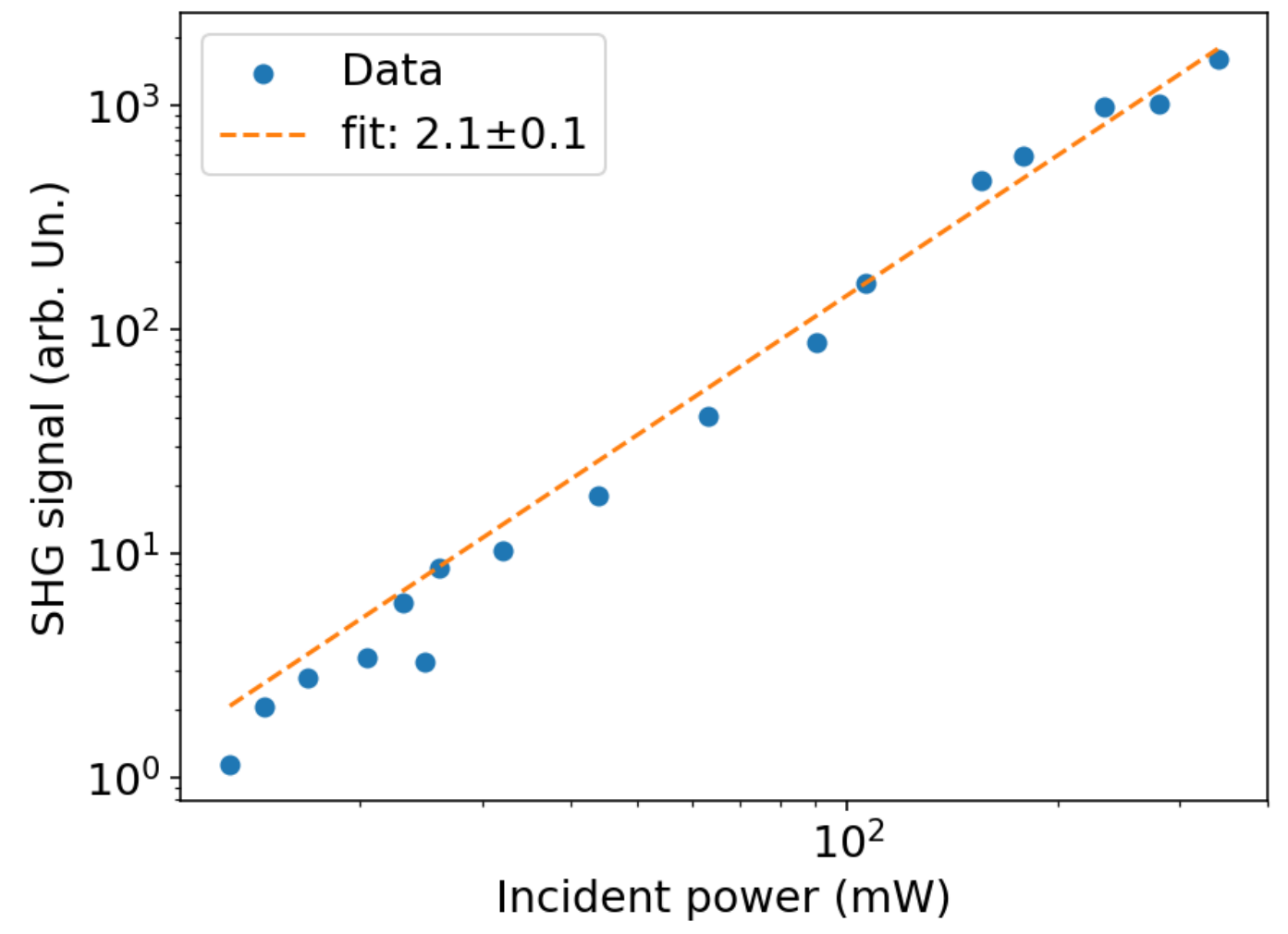}
\caption{Total SHG signal from a single shell as a function of incident power of the fundamental laser beam. The blue dashed line show a power law fit of the form $y=x^s$ yielding $s = 2.1 \pm 0.1$ }\label{figS10}
\end{figure}

\clearpage
A characteristic signature of the SHG process is the quadratic dependence of the SHG signal power on the power of the fundamental optical beam. The measured dependence for one of the larger shells is shown in Fig.~\ref{figS10}. A fit to the data yields an exponent of 2.071 $\pm$ 0.099, confirming the expected quadratic behaviour. This indicates that other light-induced effects on the liquid crystal structure, such as reorientation or photobleaching, are negligible under the applied experimental conditions.

\begin{figure}[h!]
\centering
\includegraphics[width=0.95\textwidth]{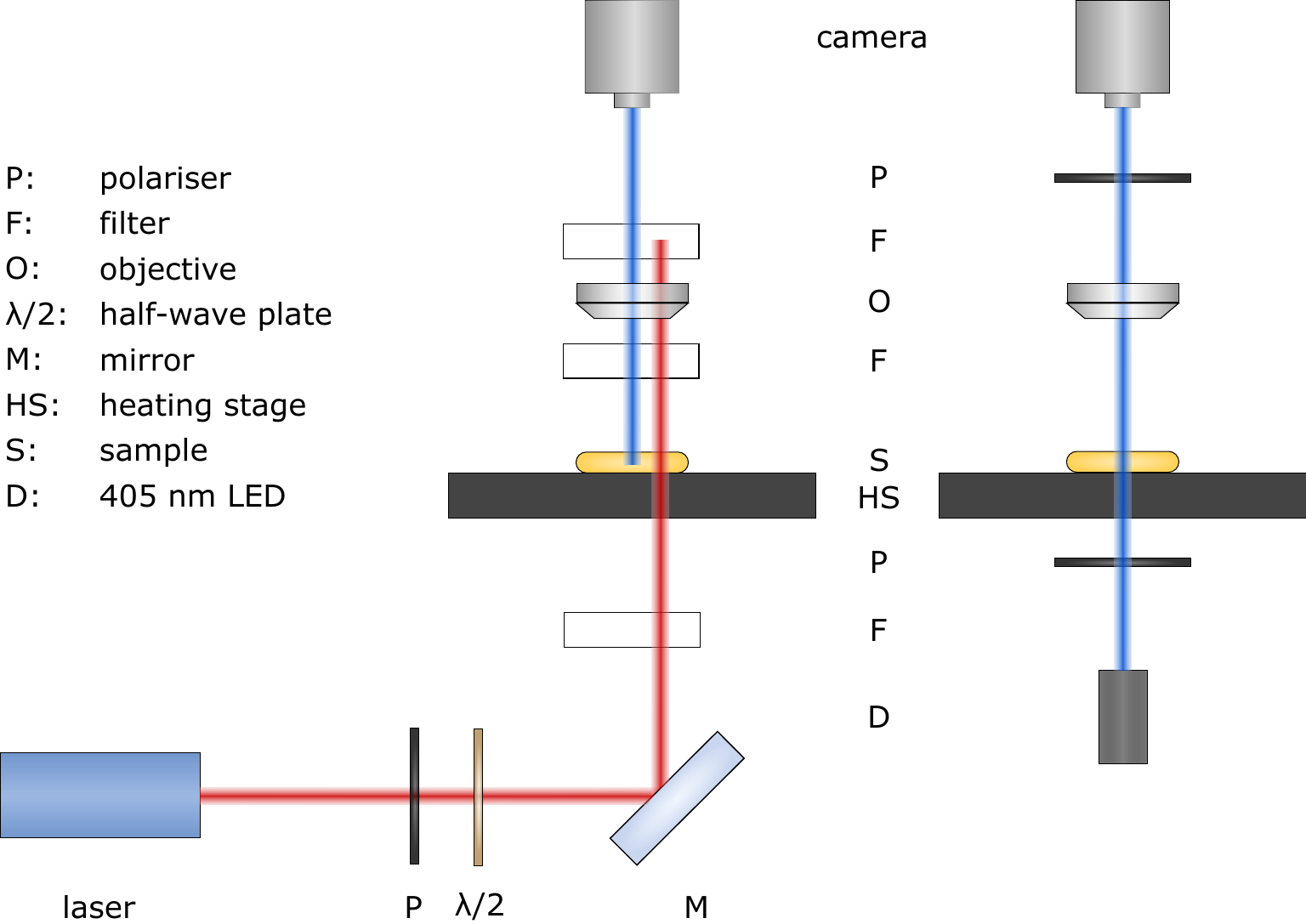}
\caption{Schematic drawing of the reconfigurable POM/SHG microscopy setup used for SHG imaging. The sample stage and imaging system (microscope objective and camera) remain fixed, while only the illumination source and selected optical components are exchanged between configurations}\label{figS11}
\end{figure}

For the interpretation of SHG microscopy results, it is highly advantageous to combine this technique with standard polarized optical microscopy (POM). To enable the use of both methods on the same selected sample, we constructed a modular, custom-made setup, as shown in Fig.~\ref{figS11}. The key advantage of this design is that both the sample mount and the imaging system remain fixed at all times, while only the illumination source and the corresponding optical components are exchanged.

\newpage
\renewcommand{\refname}{Supplementary References}
\setcounter{NAT@ctr}{41}  
\bibliographystyleSI{unsrt}       
\bibliographySI{sup-bibliography}  
\end{document}